# Modeling single- and multiple-electron resonances for electric-field-sensitive scanning probes

S. H. Tessmer[1] and I. Kuljanishvili[2]

[1]Department of Physics and Astronomy, Michigan State University, East Lansing, Michigan 48824
[2]Department of Physics and Astronomy, Northwestern University, Evanston, Illinois 60208

E-mail: tessmer@msu.edu

**Abstract**
We have developed a modeling method suitable to analyze single- and multiple-electron resonances detected by electric-field-sensitive scanning probe techniques. The method is based on basic electrostatics and a numerical boundary-element approach. The results compare well to approximate analytical expressions and experimental data.

## 1. Introduction

The ability to manipulate and probe the electrons in nanoscale systems of dopant atoms and quantum dots represents an emerging line of research. These experiments are motivated by the continued miniaturization of semiconductor devices and potential applications where single charges and spins form the functional part of the device [1-5]. Low-temperature electric-field-sensitive scanning probe methods have the potential to locally resolve electrons in these systems; such methods include scanning single-electron transistor microscopy [6, 7], charged-probe atomic force microscopy [8, 9], and subsurface charge accumulation (SCA) imaging [10, 11]. In particular, Kuljanishvili and co-workers have applied SCA imaging to probe silicon donors in an aluminum-gallium-arsenide heterostructure sample, resolving both individual electrons entering the donor layer and clusters of charge entering several donors [12].

Reference 12 briefly introduced a modeling method to simulate the capacitance-voltage curves resulting from electrons entering individual traps beneath the tip. In this paper, to fully elucidate single-electron and multiple-electron measurements, we present a detailed discussion of the electrostatic interaction. The discussion includes both analytical approximations and a numerical modeling method based on the boundary-element approach [13]. Although the discussion is motivated by SCA measurements, the approach is relevant for any capacitance-based scanning probe technique.

## 2. Subsurface Charge Accumulation method

Fig. 1(a) presents a schematic of the SCA method, which essentially measures the capacitance between the sample and a sharp metal tip. The tip is connected to a charge sensor that achieves a sensitivity of 0.01 $e/\sqrt{\text{Hz}}$ [14]. For the measurements reported here, the PtIr tip and sample were immersed in liquid helium-3 at a temperature of 290 mK. The tip's position was fixed (i.e. not scanned) at a distance of ~1 nm from the sample surface. We then monitored the AC charge $q_{tip}$ in response to a sinusoidal excitation voltage $V_{exc}$ applied to an underlying electrode, as a function of DC bias voltage $V_{tip}$. As detailed in Ref. [15], if the quantum system below the tip can accommodate additional charge, the excitation voltage causes it to resonate between the system and the underlying electrode – giving rise to an enhanced capacitance, $C = q_{tip}/V_{exc}$.

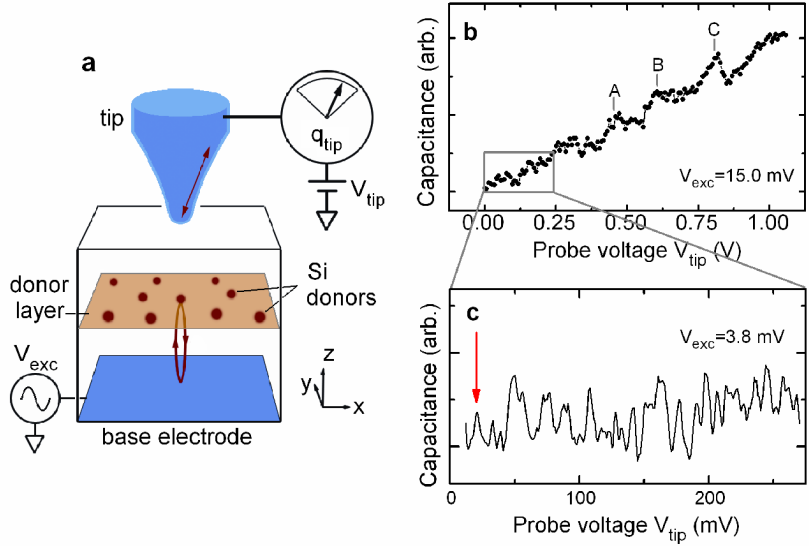

**Figure 1.** (a) Schematic of the SCA technique and key layers in the gallium-arsenide [001] heterostructure sample. An excitation voltage can cause charge to resonate between the Si donor layer and a base electrode. This results in image charge appearing on the tip. A circuit constructed from high-electron-mobility transistors is attached directly to the tip and is used to measure the charging. (b) Representative local capacitance curve measured at a single tip position, with an excitation voltage amplitude of $V_{exc}$=15mV rms. The local measurements consistently showed three broad peaks labeled A, B and C. (c) Capacitance curves acquired at the same position as part (b), but over the indicated expanded voltage range. To investigate the structure in detail, here we employed a smaller excitation amplitude of 3.8 mV rms. These data reproduced from Ref. 12.

For the measurements describe here, we employed a GaAs [001] heterostructure sample grown by molecular beam epitaxy; it contained a shallow layer of Si donors situated 20 nm above a high-mobility two-dimensional electron layer that served as an ideal base electrode for the measurement [12]. The donor plane consisted of delta-doped Si of nominal density $1.25 \times 10^{16}$ m$^{-2}$ within an $Al_{0.3}Ga_{0.7}As$ layer. The silicon atoms are confined to a plane with respect to the $z$ direction, but randomly positioned with respect to the $x$-$y$ direction (Fig. 1(a)). For the SCA measurements, the radius of interaction with the donor plane is determined approximately by the tip-donor-layer distance of 60 nm; for comparison, the tip had an apex of radius ~50 nm. Given the donor density and the expected area of interaction for the tip, we expect to be sensitive to the charging of ~150 donors.

Figs. 1(b) and 1(c) show representative capacitance-voltage curves. On the scale of 0-1 V the data show three broad peaks labeled A, B and C; the half-width-at-half-maxima (HWHM) of the peaks is roughly 50 mV. In contast, at smaller voltage scales the data consist of a series of many peaks. Although at first glance these peaks may appear to be noise, this fine structure is reproducible as long as the tip remains in the same location [12]. Moreover, individual fine peaks are consistent with single-electron charging, as discussed below. In this interpretation, larger peaks correspond to unresolved clusters of electrons.

## 3. Analysis of capacitance resonances

### 3.1 Single-electron peaks

In pioneering research in the early 90's, R. C. Ashoori and co-workers performed single-electron capacitance spectroscopy (SECS) of quantum dots [16, 17]. The experiments measured electron addition energies $\varepsilon_n$ of quantum dots in much the same way our scanned probe method measures the electron addition energies of dopant atoms. However, the original SECS work did not employ a tip; both the base electrode and the top gate were planar conducting layers. For both SECS and SCA, the movement of individual electrons between the base electrode and the quantum system, in response to a sinusoidal excitation voltage $V_{exc}$, results in an oscillation of image charge on the top gate or tip. If the DC voltage of the top-gate/tip $V_{tip}$ is slowly ramped, the resulting capacitance-versus-$V_{tip}$ curve has semi-elliptical peaks in the low-temperature limit. The HWHM of each peak is equal to 1.22 $V_{exc}$, where $V_{exc}$ is the rms value of the excitation voltage.

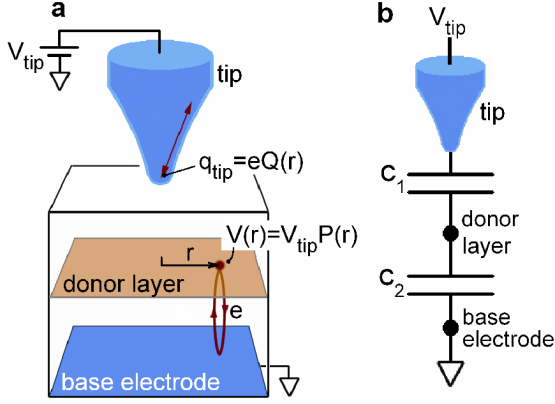

**Figure 2.** (a) Schematic of the two bell-shaped functions that determine the functional form of single-electron and multiple-electron charging. The potential function $P(r)$ gives the potential in an interior plane (donor layer) of the sample at lateral location $r$ for unit voltage applied to the tip. The charging function $Q(r)$ accounts for the charge induced on the tip due to a localized electron entering the plane at $r$. (b) Simple parallel-plate picture describing the capacitance between the sample and tip.

For SECS measurements, the voltage positions of the peaks are determined by the alignment of the base-electrode Fermi level and the addition-energy levels of the quantum system. More specifically, the peaks are centered at the gate voltages for which the chemical potential of the base electrode aligns with the addition-energy levels $\varepsilon_n$: $V_{gate} = \varepsilon_n/(\alpha e)$, where $\alpha$ is a scale factor sometimes called the voltage lever arm; $\alpha$ depends on the distances between the base electrode, quantum dot and the top gate. With regard to the amplitude of the peaks, as discussed in reference [17], an electron of charge $-e$ entering the dot results in image charge proportional to the fraction of electric flux that terminates on the top gate. This can be expressed as $q_{gate}(\text{peak}) = eC_1/(C_1+C_2)$, where $C_1$ is the mutual capacitance between the top gate and quantum dot, and $C_2$ is the mutual capacitance between the quantum dot and base electrode. For parallel-plate electrodes, $C_1/(C_1+C_2)=\alpha$.

In our work, the top electrode is a tip and hence the pattern of electric-field lines is very different from a parallel-plate picture. To analyze the measurements, we must develop a realistic model to describe the interaction between the sharp tip and planar layered sample. We find that two bell-shaped functions are key, which we refer to as the potential function $P(r)$ and the charging function $Q(r)$. $P(r)$ is essentially a position-dependent voltage lever arm. Specifically, it is the potential in the donor layer at radial position $r$ for unit voltage applied to the tip. $Q(r)$ accounts for the charge induced on the tip due to a localized electron entering the donor layer at $r$. Fig. 2(a) schematically introduces the two functions, both of which are dimensionless and maximum directly below the apex of the tip, which we define as $r=0$.

It is straightforward to describe single-electron peaks in the tip geometry in analogy to the parallel-plate SECS picture described above. Consider a charge trap that will accommodate an electron at energy $\varepsilon_0$, residing in the donor layer at position $r$. We expect the resulting capacitance-versus-$V_{tip}$ curve to be a semi-elliptical peak in the low-temperature limit. In this case, the peak will be centered at $V_{tip}=\varepsilon_0/(eP(r))$. The half-width of the peak is 1.22 $V_{exc}$, exactly as in the parallel-plate case. The non-dependence on $r$ occurs because the peak width is set by the relative magnitude of $V_{tip}$ and $V_{exc}$; both voltages must scale with the same factor of $P(r)$. However, if thermal broadening is significant, there will be $r$- dependence; the measured thermal contribution to the peak width will scale inversely with $P(r)$.

Returning to the assumption of negligible temperature, the amplitude of the single-electron peak is given by $C(\text{peak}) = q_{tip}(\text{peak})/V_{exc}$, where

$$q_{tip}(\text{peak}) = eQ(r). \quad (1)$$

Eq. (1) essentially defines the charging function $Q(r)$.

*3.2 Multiple-electron peaks*

If many charge traps are distributed within the sample, obviously, many single electron peaks can be observed. The spatial distribution of traps will lead to a spread in the measured voltages of the peaks, even if all the traps have the same addition energy. This voltage broadening is an important issue that can be considered as a limiting factor for the energy resolution of the technique.

To address the distributed-trap voltage broadening, we examine the capability of the technique to resolve a sharp resonance in the limit of an arbitrarily high density of non-interacting traps, each with the same addition energy $\varepsilon_0$. If our tip were a flat plate, then all the traps in the donor layer would charge at the same voltage, $V_{trap} = \varepsilon_0/(\alpha e)$, resulting in a sharp peak with the same functional form as a single-electron resonance. In contrast, for a realistic tip, the charging voltage must depend on the positions of the traps $r$. This leads to a characteristic capacitance peak for distributed traps $D(V_{tip})$. Here we consider the form of $D(V_{tip})$ for a realistic tip-sample geometry.

Suppose the tip voltage is significantly larger than the value for which the traps directly below the apex will charge. In this case, the charging will occur along a circle of constant $r$ (Fig. 2 (a)). In other words, a ring in the donor layer will form for which the chemical potential of the base electrode aligns with the addition-energy of the traps. The ring has an average radius $r$ for which $V_{tip}P(r) = \varepsilon_0/e$. The inner and outer radii of the ring, $r_1$ and $r_2$, are determined by $V_{tip}$, $V_{exc}$ and $P(r)$ according to the following expressions (neglecting for the moment the width of the single-electron peaks):

$$(V_{tip} - V_{exc})P(r_1) = \varepsilon_0/e,$$
$$(V_{tip} + V_{exc})P(r_2) = \varepsilon_0/e. \quad (2)$$

Next, we consider that every electron that enters the ring will induce the measured signal $q_{tip}$, as determined by $Q(r)$. Specifically, we can express the tip charging as

$$q_{tip}(V_{tip}) = \int_{r_1}^{r_2} Q(r)(2\pi r)dr. \quad (3)$$

Lastly, the resulting capacitance-versus-$V_{tip}$ curve must be convolved with the appropriate semi-elliptical function to account for the single-electron width. The procedure will yield the desired peaked function $D(V_{tip})$; the width of the peak gives the voltage resolution of the method for cases where the tip is interacting with many identical charge traps distributed below the apex.

**4. Calculating the bell functions**

*4.1 Approximate expressions*

In a thorough study, Eriksson and co-workers showed that the mutual capacitance per unit area between a scanning probe tip and a subsurface conducting layer is given by a bell-shaped Lorentzian curve,

$$c(r) = \frac{c_0}{1 + (r/w)^2}, \quad (4)$$

where $c_0$ is the capacitance per unit area of the conducting layer at $r=0$ and $w$ is the HWHM [18]. Eriksson *et al.* and Kuljanishvili *et al.* [19] showed that $w$ is equal to the depth of the layer below the exposed surface and that the accuracy of the expression holds to a few percent, if the depth of the layer is comparable to or greater than the radius of curvature of the tip's apex.

To estimate of the radial dependence of the donor-layer potential $P(r)$, we note that for our sample the donor layer is only 20 nm from the underlying 2D layer; the apex of the tip is three times farther away. Hence the contribution to the potential from the 2D layer should dominate. This contribution is proportional to the charge of the layer, which in turn is proportional to $c(r)$ as given by Eq. 4, with $w=b=80$ nm (the depth of the base electrode).

With regard to the charging function, as a first guess for $Q(r)$ we can apply a model motivated by the parallel-plate picture. In analogy to the Ashoori measurements discussed in Sec. 3.1, we approximate the function as

$$Q(r) \approx \frac{c_1(r)}{c_1(r) + c_2}, \quad (5)$$

where $c_1(r)$ is the tip-donor mutual capacitance per unit area of the donor layer, and $c_2$ is the mutual capacitance per unit area between the donor and the base electrode, which we assume to be approximately independent of donor location. This picture is shown schematically in Fig. 2(b).

To approximate $Q(r)$ we use the Lorentzian curve; specifically, for a donor of depth $a$ located at lateral position $r$, we take $c_1(r) = c(r)$, with $w=a$. This should be regarded as a rough approximation as an isolated donor is not part of a conducting layer. For the donor-to-base-electrode capacitance we use the parallel-plate expression: $c_2 = \kappa \varepsilon_0 / d$, where $\kappa$ is the dielectric constant, $\varepsilon_0$ is the free-space permittivity and $d=b-a$ is the distance between the donor layer and the base electrode. Due to the high dielectric constant of the semiconductor and the proximity of the donor layer and base electrode, we expect $c_2$ to be an order of magnitude greater than $c_1$. Hence, Eq. 5 can be further simplified to

$$Q(r) \approx \frac{c_1(r)}{c_2} = \frac{c_0 / c_2}{1 + (r/a)^2}. \quad (6)$$

So for our experiment, we expect $Q(r)$ to be roughly proportional to $c(r)$ to as given by Eq. 2, with $w=a=60$ nm.

*4.2 Numerical approach*

For a more thorough calculation of the bell functions we apply a numerical method which uses a boundary-element approach, described in detail in Reference 19. As shown schematically in Fig. 3(a), the method considers the tip and sample conductors as being composed of discrete point-like elements, and invokes image charges to account for the dielectric surface. The calculation results in a potential matrix $\hat{A}$ that can be inverted to arrive at a capacitance matrix $\hat{C}$; this large matrix gives the relationship between the voltage and charge among all the elements. In Ref. 19 the tip was modeled as a realistic cone terminated by an approximate half sphere; it was positioned 1 nm above the dielectric surface to match the experimental conditions.

To streamline the calculations presented here, the tip model is simplified; the tip is represented by a single line of points, shown schematically in Fig. 3(b). The points are spaced by 20 nm and situated such that the bottom point lays 7 nm above the dielectric surface. The total height of this line-charge tip is 400 nm. Fig. 3(c) shows an example calculation of the mutual capacitance function $c(r)$ for the tip and a conducting plane located 80 nm below the surface of an insulator with a dielectric constant of 12.5. The plane is modeled as an array of grid points with 20 nm separating adjacent points and with a total width of 800 nm.

To perform the c(r) calculation, we essentially find the charge distribution on the conducting plane required to maintain it at zero potential, with unit potential applied to the tip. As shown in Fig. 3(c), the calculated c(r) curve shows excellent agreement with the expected Lorentzian of Eq. 4, with w=80 nm. Hence we conclude that the parameters used in the boundary-element calculation are appropriate.

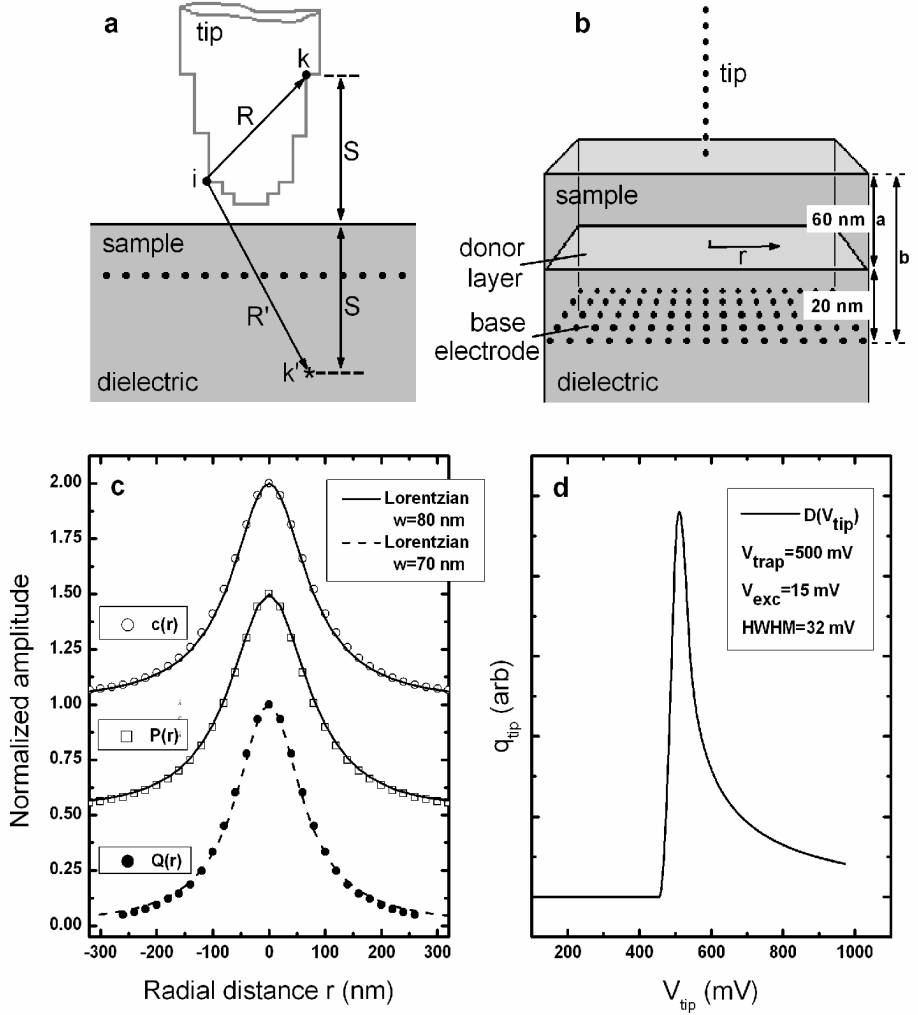

**Figure 3.** (a) Schematic example of the boundary element method [19]. Here we show the case for which both conductors $i$ and $k$ are part of the tip and $k$ is a distance $S$ from the sample dielectric surface. To calculate the potential matrix element $A_{ik}$, we consider the distance between the two conductors $R$. Moreover, to account for the modification in potential due to the dielectric layer, we consider the distance to an image charge $R'$ situated at the symmetric position with respect to the vacuum-dielectric interface. Similar considerations are used to calculate the matrix elements for the case of both conductors in the sample and for the case of one conductor in the sample and one conductor in the tip. After the complete potential matrix $\hat{A}$ is constructed for all tip and sample points, the capacitance matrix $\hat{C}$ is found by inverting $\hat{A}$. (b) The tip and sample geometry employed for the present calculations. The tip is represented by a 400 nm line of points, with each point spaced by 20 nm. It is situated such that the bottom point lays 7 nm above the dielectric surface of the sample. The sample consists of a conducting plane located at $b$=80 nm below the surface of the dielectric, with a dielectric constant of 12.5. The conducting plane is modeled as an array of grid points with 20 nm separating adjacent points and with a total width of 800 nm. Also, shown is the donor layer located at $a$=60 nm below the surface. As shown in Fig. 2(a), to calculate the bell-functions, we will consider the potential in the donor layer and the effect of introducing point charges in the layer. (c) Calculated bell-shaped curves using our boundary element method applied to the geometry shown in part (b), unless otherwise noted. The three curves are normalized so that their peak value is unity. For clarity, the middle and top curves are shifted vertically by 0.50 and 1.00, respectively, in the normalized units. (TOP) Calculation of the mutual capacitance function $c(r)$ using our method. The calculated curve is compared to the expected Lorentzian (Eq. 4) with $w$=$b$=80 nm; we see excellent agreement. (MIDDLE) Calculation of the potential function $P(r)$ compared to the expected Lorentzian curve with $w$=$b$=80 nm; we see excellent agreement. (BOTTOM) Calculation of the charging function $Q(r)$. We find that the calculation compares very well to a Lorentzian curve with $w$=70 nm, somewhat wider than the expected $w$=$a$=60 nm (Eq. 6). To save computational time while avoiding significant edge artifacts, the plotted $Q(r)$ curve is a composite of two calculations: for $r$<150 nm, we used a sample of width 400 nm with a 10 nm grid spacing; for $r$>150 nm, we used a sample of width 800 nm with a 20 nm grid spacing. (d) Calculation of the distributed-trap charging peak $D(V_{tip})$ appropriate for our tip and sample. The calculation follows from Eqs. (2) and (3), where we have used $V_{trap}$ =$\varepsilon_0/(eP(0))$=500 mV, $V_{exc}$=15 mV and the bell-functions calculated in (c) For these parameters we find a charging peak of width HWHM($D$)=32 mV, roughly double the width of a single-electron peak.

Moreover, although the line-charge tip is a highly simplified model of the actual shape, by positioning the line 7 nm above the sample surface, we achieve a very good approximation of the realistic image charge of the tip-sample system.

To calculate the potential and the charging functions, we apply our method using identical parameters as for the mutual-capacitance calculation. For $P(r)$, we consider the charge distributions on both the tip and the grounded conducting plane for unit voltage applied to the tip. These charges are then used to calculate the potential in the layer 60 nm below the surface, labeled donor layer in Fig. 3(b). To calculate $Q(r)$, we must apply the method to find the charge induced on the tip due to a point charge entering the donor layer. This is accomplished by introducing a small conducting sphere (holding unit charge) 60 nm below the surface at a series of radii *r*. The corresponding tip charge is calculated by summing over the appropriate $\hat{C}$ matrix elements [19]. For the *Q*(*r*) calculation, as the donor layer is only 20 nm from the base electrode, the 20 nm grid spacing may not be a sufficient approximation of a continuous conductor. As a check, we also performed the *Q*(*r*) calculation for a 10 nm grid spacing; we found identical functions for the two cases to a precision of about 1%.

Fig. 3(c) shows the calculated normalized *P*(*r*) and *Q*(*r*) curves. With regard to *P*(*r*), the calculation is compared to the Lorentzian curve with *w*=*b*=80 nm, which we expect to provide an approximate fit, as discussed in Sec. 4.1. Indeed, the agreement is excellent. With regard *Q*(*r*), the calculation is also compared to a Lorentzian. In this case we initially expected to achieve a good fit for *w*=*a*=60 nm, as shown by Eq. 6. However, instead the best fit was achieved for a slightly wider *w* of 70 nm, which is the dashed curve.

In summary, the agreement between the analytical expressions of Sec. 4.1 and the numerical calculations is very good, especially in light of the many approximations involved. Both approaches show that Lorentzian functions represent good approximations to the key bell-shaped curves. For the potential function, both approaches indicate a HWHM equal to the depth of the conducting layer. For the charging function, the analytical approximation predicted a HWHM equal to the donor-layer depth. However, for our sample parameters, the numerical calculation showed a HWHM equal to the average of the conducting-layer and the donor-layer depths, a disagreement of 10 nm, or about 15%.

## 5. Calculation of the distributed-trap charging peak

Given the *P*(*r*) and *Q*(*r*) curves calculated in Sec. 4, we can apply Eqs. (2) and (3) to find the *D*(*V*$_{tip}$) curve characteristic of our sample and tip.

This is a peaked function for which the width of the peak HWHM(*D*) gives the voltage resolution of the method for densely distributed charge traps. Fig. 3(d) shows the *D*(*V*$_{tip}$) appropriate for our tip and sample; for this calculation we used $V_{trap} = \varepsilon_0/(eP(0))$=500 mV and $V_{exc}$=15 mV. In general, for small excitation amplitudes, $V_{exc} \ll V_{trap}/P(0)$, we find the width depends linearly on the excitation voltage. The scale factor for the HWHM(*D*) is approximately 2.5 $V_{exc}$ for the sample/tip parameters used in our calculation, which is double the width of a single-electron peak. Moreover, we find that HWHM(*D*) has negligible dependence on the amplitudes of the bell functions *P*(*r*) and *Q*(*r*), and only a weak dependence on their widths, for variations on the scale of 10's of nm. Hence, even if we conservatively estimate that our numerical calculations of the bell curves are accurate to 20% [19], we can assert that the calculated *D*(*V*$_{tip}$) is accurate to within a few percent.

## 6. Comparison to measurements

Fig. 4(a) shows three capacitance curves taken on the same sample and at the same location as the data shown in Fig. 1. Moreover, these curves were taken at the voltage indicated by the arrow in Fig. 1(c) where a clear fine-structure peak appears in the data. This peak is a good candidate for a comparison to modeling as it is relatively well-isolated from neighboring peaks. The data are displayed to show the rms charge induced on the tip in units of the electron charge *e*, where the conversion from capacitance to charge is trivial, requiring a simple scaling by the applied excitation voltage of 3.8 mV. Fig. 4(b) shows the average of the three curves. The data are compared to a model curve which shows the semi-elliptical peak shape expected for single-electron tunneling [17]. The peak is broadened to account for the low-pass filter of the lock-in amplifier, which leads to the asymmetric shape. However, the data are not further broadened to account for temperature because this effect is much less than the 3.8 mV excitation amplitude (thermal broadening is ~$kT$=25 μV, where $T$ is temperature and $k$ is the Boltzmann constant). We see that the overall shapes of the measured and modeled curves agree reasonably well. The amplitude of the curves is about 0.075 $e$. This peak height is roughly consistent with expected captured electric flux for single-electron charging within the donor layer; for this sample we expect the factor to be approximately α=1/10 [19].

Fig. 4(c) shows data acquired on the same sample over a larger voltage range. Similar to the data of Fig. 1(b), we employed an excitation voltage of 15.0 V rms to increase the signal-to-noise ratio for the broader peaks, labeled A, B and C. To further reduce the scatter Fig. 4(c) shows an average of data acquired at three locations. Moreover, to isolate the contribution of the donor charging, we have subtracted away the background capacitance slope [12]. Given the donor density and the expected area of interaction for the tip, we expect that each of these measurement reflect the charging of ~150 donors.

In Ref. 12, we presented a donor-molecule model to account for the broad peaks. The model asserts that on average, each donor interacts with its nearest neighbor to effectively form a two-atom molecule. In this interpretation, each molecule can bind four electrons. Peaks A and B represent the average addition energies for the first and second electrons, respectively; peak C corresponds to the unresolved third and fourth electrons. As we interpret the broad peaks to arise from spatially distributed charge traps, it in reasonable to compare the peaks to the model distributed-trap curve $D(V_{tip})$, presented in Sec. 5. The solid curve of Fig. 4(c) shows the comparison (the same $D(V_{tip})$ curve presented in Fig. 3(d)). We see that the measured peak is much broader than the ideal curve predicted by distributed-trap model. This comparison was discussed briefly in Ref. 12, but not shown explicitly.

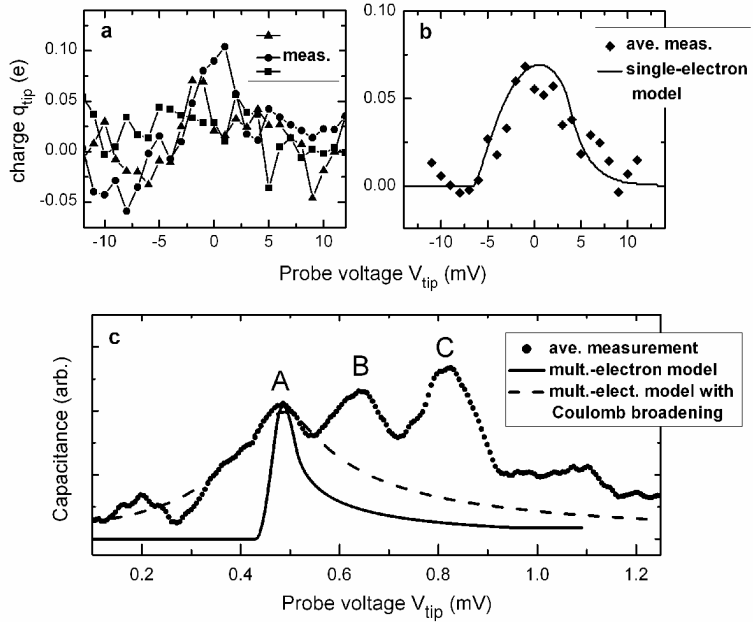

Figure 4. (a) Three curves acquired at the same location as the data of Fig. 1, near the voltage marked by the red arrow in Fig. 1(c) and with an excitation voltage of 3.8 mV. The vertical scale has been converted to charge units $q_{tip}$. (b) The average of the three measured curves shown in (a), compared to a model curve which shows the expected semi-elliptical peak shape for single-electron tunneling. (c) Capacitance measurements of the broad peaks, A, B and C. To show clearly the characteristic structure, here the measurement is the average of data acquired at three different locations. The averaging reduces the amplitude of individual single-electron peaks, which shift in voltage at different locations. However, the broader peaks, which each correspond to roughly 20 electrons tunneling at nearly the same energy, are affected little by the averaging. The excitation voltage amplitude was $V_{exc}$=15mV. Peak A is compared to two model curves. The solid curve is the $D(V_{tip})$ shown in 3(d). The dashed curve is the same $D(V_{tip})$ broadened by convolving it with a peak that accounts approximately for shifts in the addition energy due to the Coulomb interaction among the donors (the dotted curve shown in Fig. 3(d) of Ref. 12). A similar procedure can be performed for peaks B and C.

Overall, the complete donor-molecule model predicts that the characteristic capacitance curve will have three peaks; these peaks agree reasonably well with peaks A, B and C [12]. We believe that the large width of each peak arises from the fact that the molecules do not have exactly the same addition energies. In other words, with regard to peak A, unlike the assumption made in calculating the $D(V_{tip})$, the complete model does not assume that each trap has the same $\varepsilon_0$. Variations in the Coulomb interaction among the donors are likely the dominant contribution to the spread in addition energies. This is a subtle effect that is different for each of the broad peaks; essentially, the charge environment in the neighborhood of each molecule changes during the course of the measurements as charge fills the donor layer. To account for this effect for peak A, we have further broadened the $D(V_{tip})$ curve by convolving it with an approximate Coulomb-energy shift curve. The result is shown as the dashed peak in Fig. 4(c). We see that the fit is reasonable, although it neglects effects due to the overlap of the adjacent peak B.

## 7. Summary

We have developed a modeling procedure based on basic electrostatics and a boundary-element method that is suitable to analyze single- and multiple-electron resonances detected by electric-field-sensitive scanning probes. We introduce two key bell-shaped curves that are centered below the apex of the tip, the potential function and the charging function; together with the mutual capacitance curve, these functions determine the spatial and energy resolution of the methods. We find that all three bell functions are well approximated by Lorentzians of the form $[1+(r/w)^2]^{-1}$, where $r$ is the radial coordinate and $w$ is the half-width.

Our model yields curves that compare very well to approximate analytical expressions and previously-published SCA experimental data. More specifically, our modeling procedure shows that the fine structure capacitance-versus-voltage peaks observed in Ref. 12 are consistent with single-electrons entering subsurface dopants. Moreover, the broad peaks observed in the experiment are consistent with charge entering many traps distributed throughout the donor layer. We show that the increased voltage-width of these peaks can be attributed to the intrinsic width characteristic of identical but spatially distributed charge traps, as described in Sec. 3.2, and a Coulomb shift effect that further convolves the peaks, as described in Ref. 12.

## Acknowledgements

We gratefully acknowledge helpful discussions with C. Piermarocchi and T. A. Kaplan and J. F Harrison. For the experimental measurements discussed here, the sample was provided by L. N. Pfeiffer and K. W. West. This work was supported by the Michigan State Institute for Quantum Sciences and the National Science Foundation, Grant No. DMR-0305461.